# Depression of quasiparticle density of states at zero energy in $La_{1.9}Sr_{0.1}Cu_{1-x}Zn_xO_4$

C. F. Chang,[1] J.-Y. Lin,[2,*] and H. D. Yang[1]

[1]*Department of Physics, National Sun Yat-Sen University, Kaohsiung 804, Taiwan ROC*

[2]*Institute of Physics, National Chiao Tung University, Hsinchu 300, Taiwan ROC*

We have measured low-temperature specific heat $C(T, H)$ of $La_{1.9}Sr_{0.1}Cu_{1-x}Zn_xO_4$ ($x$=0, 0.01, and 0.02) both in zero and applied magnetic fields. A pronounced dip of $C/T$ below 2 K was first observed in Zn-doped samples, which is absent in the nominally clean one. If the origin of the dip in $C/T$ is electronic, the quasiparticle density of states $N(E)$ in Zn-doped samples may be depressed below a small energy scale $E_0$. The present data can be well described by the model $N(E)=N(0)+\alpha E^{1/2}$, with a non-zero $N(0)$ and positive $\alpha$. Magnetic fields depress $N(0)$ and lead to an increase in $E_0$, while leaving the energy dependence of $N(E)$ unchanged. This novel depression of $N(E)$ below $E_0$ in impurity–doped cuprates can not be reconciled with the semi-classical self-consistent approximation model. Discussions in the framework based on the non-linear sigma model field theory and other possible explanations are presented in this Letter.

**PACS**: 74.25.Bt; 74.25.Jb

The issue of quasiparticles in unconventional superconductors has attracted considerable new attention from various directions. There are now at least two questions remaining controversial. One is the extended quasiparticle states and quasiparticle transport in the mixed state [1-3], which has been intensively studied by the methods like the thermal conductivity [4,5] and specific heat [6-9]. The other is the quasiparticle density of states in the presence of impurities or disorders. Early theoretical work based on self-consistent approximation gave a general conclusion that the impurity scattering generated a residual density of states $N(E)$ at zero quasiparticle energy ($E\equiv 0$) [10-13]. Especially, the unitary scattering could lead to a nearly constant $N(E)$ near $E$=0, which was qualitatively verified by experiments [14-16]. Very recently, this scenario has been reexamined by the technique of the non-linear sigma model field theory (NLSMFT) and related numerical calculations [17-19]. This new theoretical treatment brings in a new phase transition between thermal (spin) "metals" and "insulators" in disordered $d$-wave superconductors. One of the important predictions is that $N(E)$ show a pronounced dip below a small energy scale $E_0$. In the quasiparticle localized phase, $N(E)$ is argued to vanish as $E$ or $E^2$ depending on whether the time reversal is a good symmetry or not.

In principle, the low-temperature specific heat (LTSH) $C(T)$ is a powerful probe of $N(E)$ of the quasiparticle low energy excitation. Nevertheless, previous LTSH experiments in impurity-doped (especially Zn-doped) [14,15,20] cuprates usually suffered a upturn in $C/T$ at low temperatures. This upturn hinders the investigation of the low-temperature electronic contribution in $C$, and is presumably due to either a hyperfine contribution or the local magnetic moment both of which probably are associated with defects in samples. To shed light on the issue of the low-energy quasiparticle $N(E)$, we have carefully prepared Zn- and Ni-doped $La_{1.9}Sr_{0.1}CuO_4$. These samples show no upturn in $C/T$ down to the lowest experiment temperature 0.6 K. Therefore, LTSH can be readily used to probe $N(E)$ and provide valuable information. In the following, since both Zn- and Ni-doped samples reveal the same information, the results from more intensively studied $La_{1.9}Sr_{0.1}Cu_{1-x}Zn_xO_4$ are reported.

Polycrystalline samples of $La_{1.9}Sr_{0.1}Cu_{1-x}Zn_xO_4$ with $x=0$, 0.01, and 0.02 were carefully prepared from $La_2O_3$, $SrCO_3$, and $CuO$ powder of 99.999% purity. Details of the preparation were described in [7] and references therein. The powder x-ray-diffraction patterns of all samples used in the experiments show a single T phase with no detection of impurity phases. The transition temperature $T_c$ is 33 and 14 K for $x=0$ and 0.01. The $x=0.02$ sample is not superconducting down to 2 K. The transition width (90% to 10% by the resistivity drop) of $T_c$ is 3 and 4 K for $x=0$ and 0.01 respectively, suggesting a decent homogeneity of the samples. $C(T)$ was measured from 0.6 to 8 K with a $^3$He thermal relaxation calorimeter using the heat-pulse technique. The precision of the measurement in the temperature range is about 1%. Details of the calorimeter calibrations by a standard copper sample can be found in [7]. The scatter of data in different magnetic fields is about 3% or better.

The $C(T, H=0)$ data of the samples with $x=0$, 0.01 and 0.02 are shown in Fig. 1. Compare these data, it is found that: (1) The data of the $x=0$ sample can be well fit to $C(T,0)=\gamma(0)T+C_{lattice}$, where $C_{lattice}=\beta T^3+\delta T^5$ represents the phonon contribution, with $\gamma(0)=1.54$ mJ/mol K$^2$, $\beta=0.164$ mJ/mol K$^4$, and $\delta=0.00065$ mJ/mol K$^6$. (2) Plotted as $C/T$ vs. $T^2$ in Fig. 1, the data of both $x=0.01$ and 0.02 samples are parallel to those of $x=0$ at high temperatures. Actually, fits of the data of both samples from 5 to 7 K yield almost identical $C_{lattice}$ to that of the $x=0$ sample. This is expected, as the small Zn doping should not considerably change the phonon contribution. (3) Extrapolated from the high temperature part, the intersection $\gamma$ increases significantly with increasing Zn doping. (4) Intriguingly, $C/T$ of both Zn-doped samples shows a dip at low temperatures, most evidently below 2 K, while this dip is absent in the $x=0$ sample.

The present LTSH data are usually reasoned in the context of $d$-wave pairing symmetry in cuprates [3,6-9]. For clean $d$-wave superconductors, the electronic contribution is expected to be proportional to $T^2$ at $H=0$. Indeed, this $T^2$ term has been clearly identified in overdoped $La_{1.78}Sr_{0.22}CuO_4$ [7]. The $T^2$ term leads to a downward curve at low temperatures in the plot of $C/T$ vs. $T^2$, which becomes a straight line in small magnetic fields [7], in contrast to the persistent dip of $C/T$ in strong magnetic fields observed in $La_{1.9}Sr_{0.1}Cu_{1-x}Zn_xO_4$ for $x\neq0$ (see the following paragraph). It is very likely that the absence of the $T^2$ term in the nominally clean $La_{1.9}Sr_{0.1}CuO_4$ is due to either an intrinsically small $T^2$ term or a small amount of impurities or defects in $CuO_2$ planes, as discussed in Refs [7] and [21]. The impurity scattering may cause disappearance of the $T^2$ term and generate a small but non-zero $N(0)$, which leads to at least part of $\gamma(0)T$ in LTSH. This scenario is further supported by the recent thermal conductivity measurements which show that there exists a small impurity scattering rate $\Gamma$ even in nominally pure $YBa_2Cu_3O_{6.9}$ single crystals [4]. With increasing Zn doping, pair-breaking due to larger $\Gamma$ generates larger $N(0)$ which leads to an increase in $\gamma$. In the unitary limit (which is widely assumed and supported by experimental results [4,22,23]), the electronic specific heat due to the quasiparticle states should have a linear $T$ dependence. Therefore, the dip of $C/T$ in both Zn-doped samples is certainly extraordinary. To show how peculiar the dip is, LTSH of Cu was plotted as the dash line in Fig. 1. LTSH of Cu is known to come from a constant $N(E)$ near the Fermi energy and the phonon contribution. As expected, $C/T$ of Cu shows a straight line in this temperature range.

We have also studied the magnetic field dependence of $C(T, H)$ of all these samples. For $x=0$, the increase in $\gamma$ is proportional to $H^{1/2}$ as reported in [7], consistent with the Volovik's predictions. For the $x=0.01$ sample, an increase in $\gamma$ due to applied magnetic fields was observed above 3 K. Below 3 K, however, the increase in $C/T$ due to $H$ becomes less significant as shown in Fig. 2(a). Indeed, $C/T(H=4$ T) intends to be smaller than $C/T(H=0)$ at very low temperatures. (see inset of Fig. 2(a)) This plot shows that, with increasing $H$, depression of LTSH at low temperatures becomes more significant, and the temperature $T_0$, below which the depression occurs, increases. Applied magnetic fields lead to similar effects on $C(T,H)$ of the $x=0.02$ sample as shown in Fig. 2(b). The low-temperature depression of $C/T$ due to $H$ is even more obvious since $H$ does not lead to a significant increase in $\gamma$ for the non-superconducting sample in contrast to the $x=0.01$ one.

One of the possible origins of the dip in $C/T$ of the Zn-doped samples is depression of $N(E)$ below $E_0$. It is noted that this possibility can not be reconciled with previous theoretical works based on semi-classical self-consistent approximation, which in general leads to a constant $N(E)$ [10-12]. Very recent works of the non-linear sigma model field theory and the related numerical calculations on the dirty high-temperature superconductors, however, reveal a vanishing $N(E)$ among other things like the localization of the quasiparticles [13,16-18]. In this framework, quasiparticles are always localized in the two-dimensional dirty $d$-wave system, if quasiparticle interactions can be ignored. Furthermore, $N(E)$ vanishes as $E$ or $E^2$ depending on whether the time reversal is a good symmetry or not.

To compare the data with the present theory, $C(T,H)$ below 2 K has been analyzed based on the model $N(E)=N(0)+\alpha E^\nu$. This analysis leads to a non-zero $N(0)$ which is further depressed by $H$. Meanwhile, it is found that $\nu=1/2$ gives a better fit than $\nu=1$, although both values of $\nu$ qualitatively describe the data. The fitting results for $\nu=1/2$ are shown as the solid lines in Fig. 3. This non-zero $N(0)$ resulting from LTSH is consistent with the results of thermal conductivity measurements in Zn-doped YBCO [4]. Interestingly, the case of a non-zero $N(0)$ with $\nu=1/2$ coincides with the case of the "thermal metal" in NLSMFT [19]. If this is the case, it is not clear whether interactions between quasiparticles are the cause of delocalization.

However, this assumption is not implausible since cuprates are known to be systems of strong correlation. Whatever the underlying mechanism is, $N(E)=N(0)+\alpha E^\nu$ rather than $N(E)=N(0)$ was clearly observed, perhaps indicating a crossover from diffusive to localization regime of the quasiparticles in $La_{1.9}Sr_{0.1}Cu_{1-x}Zn_xO_4$. In this scenario, $C/T(T=0)$ is proportional to $N(0)$. The fits yield $C/T(T=0)=4.92\pm0.03$, $4.90\pm0.05$, and $4.80\pm0.03$ mJ/mol K$^2$ for $H=0$, 1, and 4 T, respectively. These results suggest that $N(0)$ is depressed by $H$. This depression effect of $H$ on $N(0)$ can also be seen in Fig. 2(b) even without fitting. We summarize $N(E)$ of $La_{1.9}Sr_{0.1}Cu_{0.99}Zn_{0.01}O_4$ suggested from LTSH by Fig. 4. $N(E)$ dips to a non-zero $N(0)$ below the energy scale $E_0/k\sim 2$ K. In the presence of magnetic fields, $N(E)$ above $E_0$ increases due to the Doppler shift proposed by Volovik [3]. More importantly, magnetic fields raise $E_0$ and further depress $N(0)$, while the energy dependence of $N(E)$ remains unchanged.

We have measured $C(T, H)$ of $La_{1.9}Sr_{0.1}Cu_{1-x}Ni_xO_4$ and $La_{1.78}Sr_{0.22}Cu_{1-x}Ni_xO_4$. While the results of $La_{1.9}Sr_{0.1}Cu_{1-x}Ni_xO_4$ are the same as those of $La_{1.9}Sr_{0.1}Cu_{1-x}Zn_xO_4$ reported above, $C/T$ of $La_{1.78}Sr_{0.22}Cu_{1-x}Ni_xO_4$ shows no dip. This is probably because the underdoped samples are more two-dimensional than the overdoped ones. On the other hand, it is not theoretically clear whether depression of $N(E)$ would still take place when a $d$-wave superconductor is impurity-doped to become non-superconducting[24]. Finally, there are other possible explanations about the depression of $C/T$ (or $N(E)$) to be discussed. Since the dip of $C/T$ is influenced by magnetic fields, the explanation of optical phonons can be ruled out [25]. By the same token, the depression of $N(E)$ due to Coulomb interactions is unlikely to be the case[26]. At low temperatures, a second superconducting phase, induced by impurity doping, could possibly develop a small gap and lead to depression of $C/T$. However, this second superconducting phase, if it exists, should be suppressed by applied magnetic fields rather than be enhanced [27].

Zn substitution may disturb the magnetic correlation in $CuO_2$ planes and induce the local moment on the four neighboring Cu sites [28]. Assuming that the local moment is spin-1/2, it could contribute to LTSH according to the Schottky form. However, it is found that the magnetic contribution of $4x$ paramagnetic centers (PC's) is too large to be incorporated into LTSH data of the Zn-doped samples. Very recent studies actually indicated an overwhelming screen of the Schottky anomaly in LTSH and a partial screening of the induced moment in susceptibility measurements for Zn-doped cuprates [20,29]. If the dip in $C/T$ were attributed to the magnetic origin, an effective field of 4.2 T would be required at zero applied field for $x=0.01$ sample. Meanwhile, the fit would lead to a PC concentration of 0.02% rather than 4%. Though this effective field is not particularly large, it has never been observed in samples with the same or larger order of PC concentration [7,20,21]. The analysis for $x=0.02$ data in this context would lead to the same implausible results, too. To further examine the possibility of the magnetic interpretation, we have fit data by $C(T,H)=\gamma(H)T+\beta T^3+\delta T^5+ nC_{Schottky}(T,H+H_0)$ where $\gamma(H)T$ is the electronic contribution, $\beta T^3+\delta T^5$ are the phonon contribution, the last one is the Schottky anomaly, and $H_0$ is the effective field in the sample. These fits result in a nonplusing *decrease* in $\gamma$ with increasing $H$ for both samples. Furthermore, the curvature of the fitting curves does not match that of data at low temperatures, especially for large $H$. Thus, the magnetic interpretation is less plausible though it can not be totally eliminated as a possible explanation.

In summary, we have carefully measured $C(T, H)$ of $La_{1.9}Sr_{0.1}Cu_{1-x}Zn_xO_4$ and found an intriguing depression of $C/T$ at low temperatures in Zn-doped samples. If the origin of this peculiar phenomenon is electronic, the present LTSH data suggest a dip of $N(E)$ at zero energy. This novel quasiparticle density of states $N(E)$ is sensitive to the applied magnetic fields as shown in Fig. 4. It is hoped that this Letter is one of the first experimental efforts toward to a full understanding of the quasiparticle states in impurity-doped cuprates.

We are grateful to Hsiu-Hau Lin, S. Vishveshwara, M. P. A. Fisher and N. E. Phillips for indispensable discussions. This work was supported by National Science Council of Republic of China under Contract Nos. NSC89-2112-M-110-008 and NSC89-2112-M-009-007.

**Figure Captions**

Fig. 1. $C/T$ vs. $T^2$ of La$_{1.9}$Sr$_{0.1}$Cu$_{1-x}$Zn$_x$O$_4$. The specific heat of copper is included as the dash line to contrast with the low-temperature dips in both Zn-doped samples.

Fig. 2. (a) The effects of magnetic fields on $C(T, H)$ of the $x$=0.01 sample. Inset: the enlargement of the low-temperature part. (b) $C(T, H)$ of the $x$=0.02 sample.

Fig. 3. The LTSH data at low temperatures can be well described based on the model $N(E)=N(0)+\alpha E^{1/2}$ (see text).

Fig. 4. The proposed quasiparticle $N(E)$ of the impurity-doped cuprates based on LTSH data. The scale of $E_0$ is exaggerated. In reality, $E_0/k$~2 or 3 K in La$_{1.9}$Sr$_{0.1}$Cu$_{1-x}$Zn$_x$O$_4$. $E_0/E_1$ is order of 10 in the $x$=0.01 sample, where $E_1$ is the energy scale above which $N(E)$ is no longer constant. The magnitude of $E_1$ is estimated from the results in [10] by $T_c(x$=0.01$)/T_c(x$=0$)$=0.42.

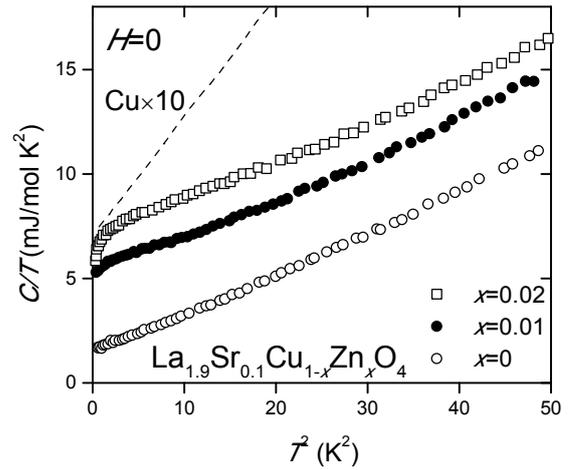

Fig. 1

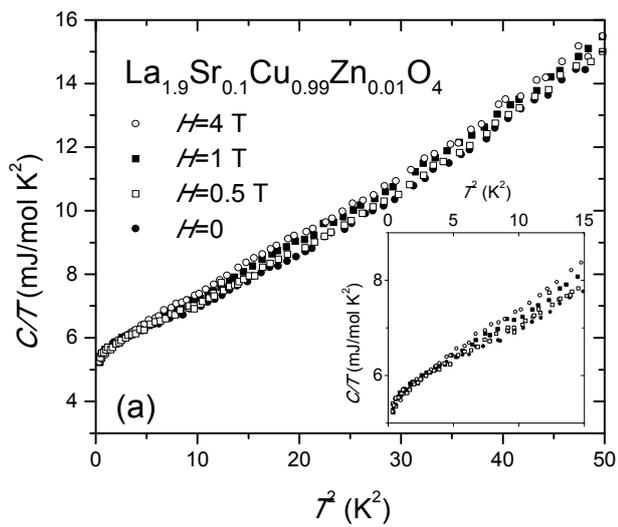

Fig. 2(a)

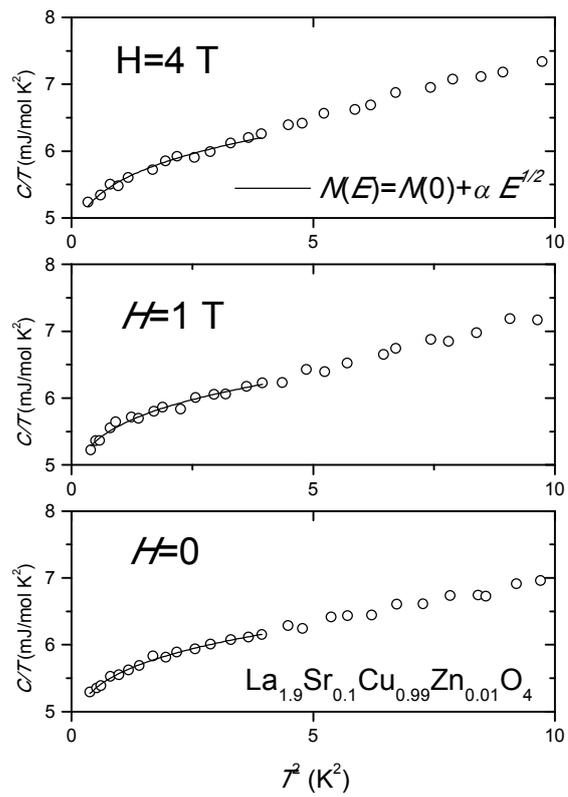

Fig. 3

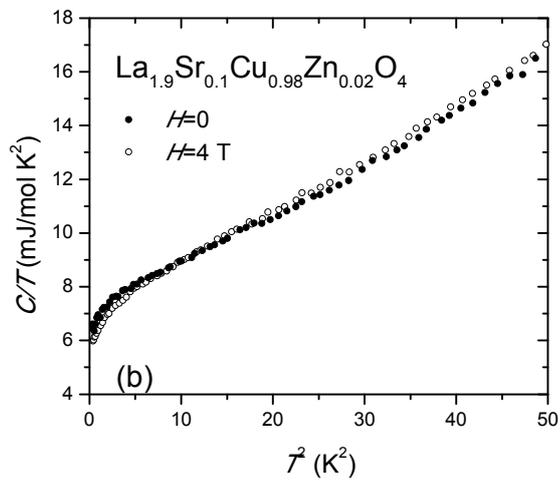

Fig. 2(b)

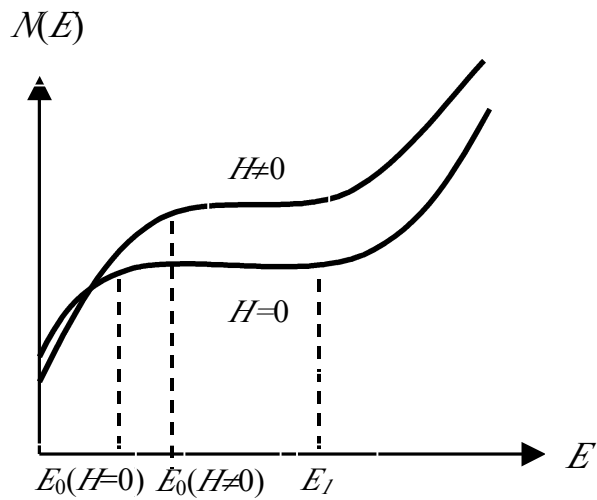

Fig. 4